\documentclass[%
 reprint,
 amsmath,amssymb,
 aps,prd,
]{revtex4-2}

\usepackage{graphicx}% Include figure files
\usepackage{dcolumn}% Align table columns on decimal point
\usepackage{bm}% bold math
\usepackage{hyperref}% add hypertext capabilities
\usepackage[usenames,dvipsnames]{xcolor}
\usepackage{amsmath}
\usepackage{amssymb}
\usepackage{orcidlink}
\usepackage[capitalize]{cleveref}
\usepackage{float}

\crefname{section}{Sec.}{Secs.}
\Crefname{section}{Sec.}{Secs.}

\newcommand{\gw}{\mathrm{GW}}
\newcommand{\Msun}{M_\odot}
\newcommand{\hp}{h_+}
\newcommand{\hc}{h_\times}

\begin{document}

\preprint{APS/123-QED}

\title{Uncertainty in predicting the stochastic gravitational wave background from compact binary coalescences}

\author{Michael Ebersold \,\orcidlink{0000-0003-4631-1771}
}

%\affiliation{Laboratoire d'Annecy de Physique des Particules, CNRS, 9 Chemin de Bellevue, 74941 Annecy, France}
\affiliation{Physik-Institut, University of Zurich, Winterthurerstrasse 190, 8057 Zurich, Switzerland}

\email{michael.ebersold@physik.uzh.ch} 

\author{Tania Regimbau}

\affiliation{Univ. Savoie Mont Blanc, CNRS, Laboratoire d’Annecy de Physique des Particules - IN2P3, F-74000 Annecy, France}

\date{\today}

\begin{abstract}

The stochastic gravitational wave background from compact binary coalescences is expected to be the first detectable stochastic signal via cross-correlation searches with terrestrial detectors. It encodes the cumulative merger history of stellar-mass binaries across cosmic time, offering a unique probe of the high-redshift Universe. However, predicting the background spectrum is challenging due to numerous modeling choices, each with distinct uncertainties.
In this work, we present a comprehensive forecast of the astrophysical gravitational wave background from binary black holes, binary neutron stars, and neutron star--black hole systems. We systematically assess the impact of uncertainties in population properties, waveform features, and the modeling of the merger rate evolution.
By combining all uncertainties, we derive credible bands for the background spectrum, spanning approximately an order of magnitude in the fractional energy density. These results provide thorough predictions to facilitate the interpretation of current upper limits and future detections.
\end{abstract}

\maketitle

\section{Introduction}

The gravitational wave observatories LIGO~\cite{Aasi:2015}, Virgo~\cite{Acernese:2015}, and KAGRA~\cite{KAGRA:2020} are gradually revealing the population of merging stellar-mass compact binaries in the Universe~\cite{GW150914:2016,GW170817:2017,LIGO_GWTC-1:2019,LIGO_GWTC-2:2020,GWTC-3:2021}. By directly observing hundreds of binary black hole (BBH) mergers, these detectors have provided unprecedented insights into the population of compact objects~\cite{LIGO_O2pop:2019,LIGO_O3apop:2020,GWTC3_population}. 
While individual detections provide detailed information about the local population, a vast number of mergers remain undetected, either because they occur at high redshifts or because their signals fall below the sensitivity threshold of current instruments. The superposition of these unresolved compact binary coalescences (CBCs) produces a persistent stochastic gravitational wave background (SGWB)~\cite{Phinney:2001,Regimbau:2022}, which encodes the integrated history of binary mergers across cosmic time.

The search for a SGWB of astrophysical origin complements the direct observations of individual events, as it provides access to the high-redshift Universe~\cite{GW150914Stoch:2016,GW170817Stoch:2018,LIGO-O1-stoch:2016,LIGO-O2-stoch:2019,LVK-isotropic-o3:2021,Callister:2020}. 
While this work focuses on the SGWB produced by CBCs, other astrophysical and cosmological sources may also contribute to the background. These include core-collapse supernovae~\cite{Sandick:2006,Buonanno:2004,Zhu:2010}, rotating neutron stars~\cite{Ferrari:1998,Regimbau:2001,Rosado:2012,Lasky:2013}, primordial black holes~\cite{Carr:2020,Mukherjee:2021,Boybeyi:2024}, and various early-Universe phenomena~\cite{Kosowsky:1992,Hindmarsh:2013,Hindmarsh:2015,Caprini:2018,Caprini:2024}.
Forecasts of the astrophysical gravitational wave background (AGWB) from CBCs have evolved significantly over time. Early predictions, prior to the first gravitational wave (GW) detections, relied on simple models of binary populations~\cite{Regimbau:2005,Regimbau:2007,Regimbau:2011,Marassi:2011,Zhu:2011,Rosado:2011,Wu:2012,Zhu:2013}. More recent studies incorporate population synthesis models~\cite{Kowalska:2015,Perigois:2020,Perigois:2021} or observational data into the projections~\cite{LVK-isotropic-o3:2021,GWTC3_population,Bellie:2023,Lehoucq:2023,Renzini:2024}. 
Despite these advances, predicting the AGWB spectrum remains challenging due to the numerous astrophysical inputs involved, many of which carry substantial uncertainties.
These include the mass and spin distribution of the compact binary population, as well as the redshift-dependent merger rate.
In particular, modeling the merger rate history is complicated by factors that are difficult to capture with simple prescriptions and vary across different formation channels~\cite{Mapelli:2017,Rodriguez:2018,Baibhav:2019,Perigois:2021,Kou:2024}. Among these are the delay time between binary formation and merger, which is sensitive to the underlying stellar evolution and binary interaction physics, and the metallicity of the progenitor environment, which affects both the formation efficiency and the physical properties of the resulting compact objects~\cite{Dominik:2013,Dominik:2014,Nakazato:2016,Dvorkin:2016,Mapelli:2018,Mapelli:2019,Turbang:2023}. 
This work focuses on two key aspects of the AGWB from CBCs: (1) accurately modeling the AGWB while properly accounting for the various uncertainties involved in its calculation, and (2) systematically analyzing the impact of individual modeling assumptions on the resulting background spectrum.
Our aim is to provide a comprehensive understanding of how different physical inputs and modeling choices shape the AGWB spectrum, and to quantify the credible range within which the background is expected to lie.

We begin by introducing a fiducial model for the AGWB from three classes of CBCs: BBH, binary neutron star (BNS), and neutron star--black hole (NSBH) systems. These models reflect the current best estimates of the local binary population, as characterized in Ref.~\cite{GWTC3_population}, and adopt the merger rate evolution used in the forecast of Ref.~\cite{LVK-isotropic-o3:2021}.
To compute the AGWB spectrum, we evaluate the fractional energy density via Monte Carlo integration over the relevant source parameters, employing state-of-the-art waveform models.
We then investigate how the uncertainties in the mass and spin distribution propagate into uncertainties in the AGWB spectrum. We also show how different population models influence the spectral shape and amplitude of the background.
Additionally, we examine the impact of waveform features beyond the leading-order quadrupole approximation, including higher-order GW modes in BBH mergers and tidal effects in BNS systems, both of which can significantly alter the spectral features of the background.

An important ingredient in modeling the AGWB from compact binaries is the history of their merger rates, extending to redshifts that are not directly accessible with the current generation of GW detectors, yet still influence the AGWB spectrum.
A common approach to modeling this evolution involves convolving the cosmic star formation rate with a delay-time distribution that accounts for the time between binary formation and merger.
For BBHs, this model often includes a metallicity threshold, allowing BBH formation only in environments below a certain metallicity. Each of these modeling components, star formation history, delay-time distribution, and metallicity dependence, introduces substantial uncertainties that propagate into the predicted AGWB spectrum. 
Here, we systematically investigate the individual and combined effects of these uncertainties. In particular, we explore a range of delay-time distributions, varying both the power-law index and the minimum delay time. We also examine different assumptions regarding the metallicity threshold for BBH formation, focusing on scenarios with varying degrees of dispersion around the mean metallicity. 
By sampling a broad parameter space, including local rates, delay-time distributions, and metallicity thresholds, we construct an ensemble of merger histories to quantify the combined uncertainty.
Our analysis demonstrates that these factors can significantly impact the merger rate history and, consequently, the AGWB spectrum. 

Finally, we combine the ensemble of merger rate histories with the uncertainties in the population models to construct credible bands for the AGWB spectrum from BBH, BNS, and NSBH systems. These bands represent the range of plausible background amplitudes and span approximately an order of magnitude in $\Omega_\gw (f)$. Compared to most estimates in the literature, our bands are notably broader, mainly due to the inclusion of uncertainties in the merger-rate modeling. 
With the upcoming observing runs of the LVK collaboration, cross-correlation searches are expected to start directly constraining the most optimistic predictions for the AGWB. 

This paper is structured as follows: In~\cref{sec:fiducial}, we introduce the fiducial model of the AGWB from compact binaries, which serves as a reference for evaluating the impact of different modeling choices. \Cref{sec:uncertainty-population} discusses uncertainties arising from population models, including modeling choices such as incorporating higher-order waveform modes in computing the BBH background and tidal effects for the BNS background. In~\cref{sec:uncertainty-mergerrates}, we analyze the individual and combined uncertainties associated with modeling the merger rate history. In~\cref{sec:combined-uncertainty}, we combine all uncertainties and present the final credible bands for the AGWB from the three CBC classes. Finally, we conclude with a discussion and outlook in~\cref{sec:conclusion}.

\section{Fiducial Model of the astrophysical background from CBC}
\label{sec:fiducial}

A Gaussian, isotropic, unpolarized, and stationary GWB is characterized by the fractional energy density spectrum~\cite{Allen:1996,Romano:2016},
\begin{align}
    \Omega_\gw (f) = \frac{f}{\rho_c} \, \frac{d\rho_\gw}{df}\,,
\end{align}
where $d\rho_\gw$ denotes the GW energy density contained in the frequency interval $f + df$, $\rho_c = \frac{3 H_0^2 c^2}{8 \pi G}$ is the critical energy density required today to close the Universe, and $H_0$ is the Hubble constant. In general, the stochastic GWB from an astrophysical population can be computed using the following expression~\cite{Phinney:2001,Regimbau:2011}:
\begin{align} \label{eq:OmegaGW}
    \Omega_\gw(f)= f \int p(\theta_i) d \theta_i \int_{0}^{z_{\max}} d z \frac{R_m (\theta_i, z) \; \frac{d E_\gw\left(\theta_i, f_s\right)}{d f_s}}{\rho_c\,(1+z) H(z)} \,,
\end{align}
where the second integral extends from the present day to a maximum redshift $z_{\max}$ corresponding to the onset of binary mergers in the early universe.
Here, $p(\theta_i)$ denotes the probability distributions of the source parameters $\theta_i$, $R_m(\theta_i, z)$ is the merger rate density in the source frame, and $\frac{d E_\gw\left(\theta_i, f_s\right)}{d f_s}$ is the energy spectrum emitted by a single source, evaluated in the source frame at $f_s = f(1+z)$. The Hubble parameter $H(z) = H_0 \sqrt{\Omega_M (1+z)^3 + \Omega_\Lambda}$ encodes the cosmological expansion, assuming a flat $\Lambda$ cold dark matter ($\Lambda$CDM) cosmology with Hubble constant $H_0$, matter density $\Omega_M$, and dark energy density $\Omega_\Lambda$. 

In this section we introduce fiducial models for the GWB generated from different CBCs, which serve as a baseline for comparing different sources of uncertainty. These fiducial models follow similar assumptions as in Ref.~\cite{LVK-isotropic-o3:2021}, enabling straightforward comparison.

\subsection{Fiducial binary populations}
\label{sec:binarypop}

For the BBH population, we adopt the best-fit mass distribution inferred from GWTC-3 based on the confident detections with an inverse false alarm rate greater than one year. While our analysis is based on GWTC-3 population inferences, we note that GWTC-4~\cite{GWTC-4-results:2025,GWTC-4-population:2025} has since been released, and its potential impact on our results is discussed in~\cref{sec:conclusion}. Specifically, we use the preferred power-law plus peak (PLP) model described in the Appendix of Ref.~\cite{GWTC3_population}. In this model, the primary mass distribution follows a power law with spectral index $a$ between $m_\mathrm{min}$ and $m_\mathrm{max}$ with an additional Gaussian peak with mean $m_\mu$ and width $m_\sigma$. The mixing fraction $\lambda_p$ determines the relative contribution of the Gaussian peak and the power-law component. Additionally, the parameter $\delta_m$ governs the smoothing at the lower end of the mass spectrum. The mass ratio $q=m_2/m_1$ distribution follows a power law with spectral index $\beta_q$. 
The hyperparameters defining the fiducial PLP model are set to $a = 3.5$, $m_\mathrm{min} = 5.1 \,\Msun$, $m_\mathrm{max} = 88 \, \Msun$, $\lambda_p = 0.038$, $m_\mu = 34 \,\Msun$, $m_\sigma = 4.6 \,\Msun$, $\delta_m = 5.0 \, \Msun$ and $\beta_q = 1.1$, based on the median values from the population inference presented in Ref.~\cite{GWTC3_population}.
Given that the observed BBH population exhibits low effective inspiral spins~\cite{GWTC3_population,Miller:2020}, we assume nonspinning BBHs in our fiducial model. Nonetheless, in~\cref{sec:spins} we explore the impact of alternative spin distributions.

Due to the small number of BNS events compared to BBH events, the BNS mass distribution remains poorly constrained. For our fiducial model, we follow Ref.~\cite{LVK-isotropic-o3:2021} and assume a uniform distribution for both component masses, $m_1$ and $m_2$, in the range $1.0\,\Msun$ to $2.5\, \Msun$. Tidal effects due to the matter in neutron stars are neglected for the fiducial model. Additionally, we assume nonspinning BNS systems, in agreement with expectations from Galactic BNS spin measurements~\cite{Tauris:2017}.
The mass distribution of NSBH is similarly uncertain. Following Ref.~\cite{GWTC3_population}, we adopt a uniform distribution of neutron star masses between $1 \, \Msun$  and $2.5\, \Msun$, and a logarithmically uniform distribution for black hole masses between $5 \, \Msun$ and $50 \,\Msun$.

\subsection{Binary merger rates}

For all fiducial CBC backgrounds, we model the source-frame merger rate $R_m(z)$ as following the star formation rate (SFR) using the parametrization of Ref.~\cite{Madau:2014},
\begin{align} \label{eq:SFR}
    R_f (z) \propto \frac{(1+z)^{\gamma}}{1+\left(\frac{1+z}{1+z_\mathrm{peak}}\right)^{\kappa}} \,,
\end{align}
where we set the values $\gamma =2.6$, $z_\mathrm{peak} = 2.2$, and $\kappa = 6.2$, according to Madau-Fragos~\cite{Madau:2016}. With redshift the rate first increases with the power-law index $\gamma$, peaks at $z_\mathrm{peak}$ and decreases again for high redshifts with negative power-law index $\kappa$. 
Although Ref.~\cite{LVK-isotropic-o3:2021} employed a different SFR parametrization following Ref.~\cite{Springel:2002,Vangioni:2014}, we demonstrate later that the choice of SFR model has only a minor impact on the resulting GWB.

To account for the time between formation and merger of compact binaries, we incorporate a time delay $t_d$, which reflects the evolutionary timescale from the birth of massive stars to the formation of the compact objects and their eventual merger. The time delay connecting the redshift at formation $z_f$ and merger $z$ can be expressed in terms of the cosmological lookback times
\begin{align}
    t_d = t_c (z_f) - t_c (z) \,,
\end{align}
where
\begin{align}
    t_c (z) = \int^z_0 \frac{dz'}{(1+z') H(z')} \,.    
\end{align}
The merger rate $R_m(z)$ is then computed as a convolution of the SFR $R_f(z_f)$ with the probability distribution of time delays $p(t_d)$,
\begin{align}
    R_m (z) = \int^{t_{d, \mathrm{max}}}_{t_{d, \mathrm{min}}} d t_d \, R_f (z_f (z, t_d))\, p (t_d)\,,
    \label{eq:Rmz}
\end{align}
and normalized to the local merger rate inferred from GW observations~\cite{GWTC3_population}. 
For BBHs modeled with the PLP mass distribution, the rate is best constrained at a redshift $z=0.2$ with a value of $28.3 \,\mathrm{Gpc^{-3} \mathrm{yr}^{-1}}$. For the fiducial mass distribution of BNS and NSBH systems, the reported rates at $z = 0$ are $105.5 \,\mathrm{Gpc^{-3} \mathrm{yr}^{-1}}$ and $32.0 \, \mathrm{Gpc^{-3} \mathrm{yr}^{-1}}$, respectively.
The time-delay distribution $p(t_d)$ is typically assumed to be log-uniform, with a maximum delay equal to the Hubble time and a minimal delay of a few Myr. In our fiducial model, we adopt a minimum delay of 50 Myr for BBHs and 20 Myr for both BNS and NSBH systems.
Additionally, when computing the BBH merger rate, we weight the SFR at the formation redshift $z_f$ by the fraction of stellar formation occurring below a metallicity threshold $Z < Z_\mathrm{th}$. This accounts for the enhanced efficiency of BBH formation in low-metallicity environments~\cite{Mapelli:2019,Chruslinska:2018}. The metallicity-dependent efficiency is modeled following Ref.~\cite{Langer:2005}, as also adopted in Refs.~\cite{LVK-isotropic-o3:2021,GWTC3_population,Turbang:2023}, 
\begin{align} \label{eq:gammetal}
    \varepsilon_Z (z) = \hat\Gamma \left( 0.84, (Z_\mathrm{th}/Z_\odot)^2 \, 10^{0.3 z} \right) \,,
\end{align}
where $\hat\Gamma$ denotes the incomplete gamma function. For the fiducial BBH model, we assume $Z_\mathrm{th} = 0.1\, Z_\odot$, and different thresholds and alternative metallicity models are investigated in later sections.

\subsection{CBC background}

The energy density spectrum emitted by a single CBC source is given by~\cite{Phinney:2001}
\begin{align}
    \frac{dE_\gw(f_s)}{df_s} = 4 \pi r^2 \frac{\pi c^3}{2 G} f_s^2 \left[ \hp^2(f_s) + \hc^2(f_s) \right] \,,
\end{align}
where $\hp(f)$ and $\hc(f)$ are the plus and cross polarizations of the gravitational waveform in the frequency domain and $r$ is the distance at which the waveform is evaluated. Note that the distance dependence cancels out, as $\hp\,, \hc \sim r^{-1}$.
With all ingredients in place, we compute the GWB for the three source classes using~\cref{eq:OmegaGW}. Since this involves computing a complicated high-dimensional integral, we employ a Monte Carlo technique~\cite{Metropolis:1949}. For each variable parameter, we draw a large number $N$ of samples from their respective probability distributions and estimate the GWB via
\begin{align}
    \Omega_\gw (f) = \frac{1}{N} \frac{f \, R_n}{\rho_c} \sum^N_{k=1}  \frac{dE_\gw (f_s,\theta_i^k)}{df_s} \frac{1}{(1+z_k) H(z_k)}\,,
\end{align}
where $R_n$ is the normalization of the merger rate $R_m(z)$, which is used to construct the redshift probability distribution $p(z)$. For the fiducial CBC background, the variable parameters include the primary mass $m_1$, secondary mass $m_2$ (or equivalently the mass ratio $q = m_2 / m_1$), redshift $z$, reference phase $\phi_0$ drawn uniformly from $[0,2\pi]$, and inclination angle $\iota$, with $\cos \iota$ sampled uniformly from $[-1,1]$. 
For BBH and NSBH systems, we use the \texttt{IMRPhenomXAS}~\cite{Pratten:2020a} waveform model to compute the frequency-domain waveform polarizations. For BNS systems, we employ the \texttt{IMRPhenomXAS\_NRTidalv3} waveform model~\cite{Abac:2023,Colleoni:2023,Dietrich:2019}, which tapers the merger-ringdown portion of the baseline \texttt{IMRPhenomXAS} model and includes the option to incorporate tidal effects.

\begin{figure}[t]
    \centering
    \includegraphics[width=0.48\textwidth]{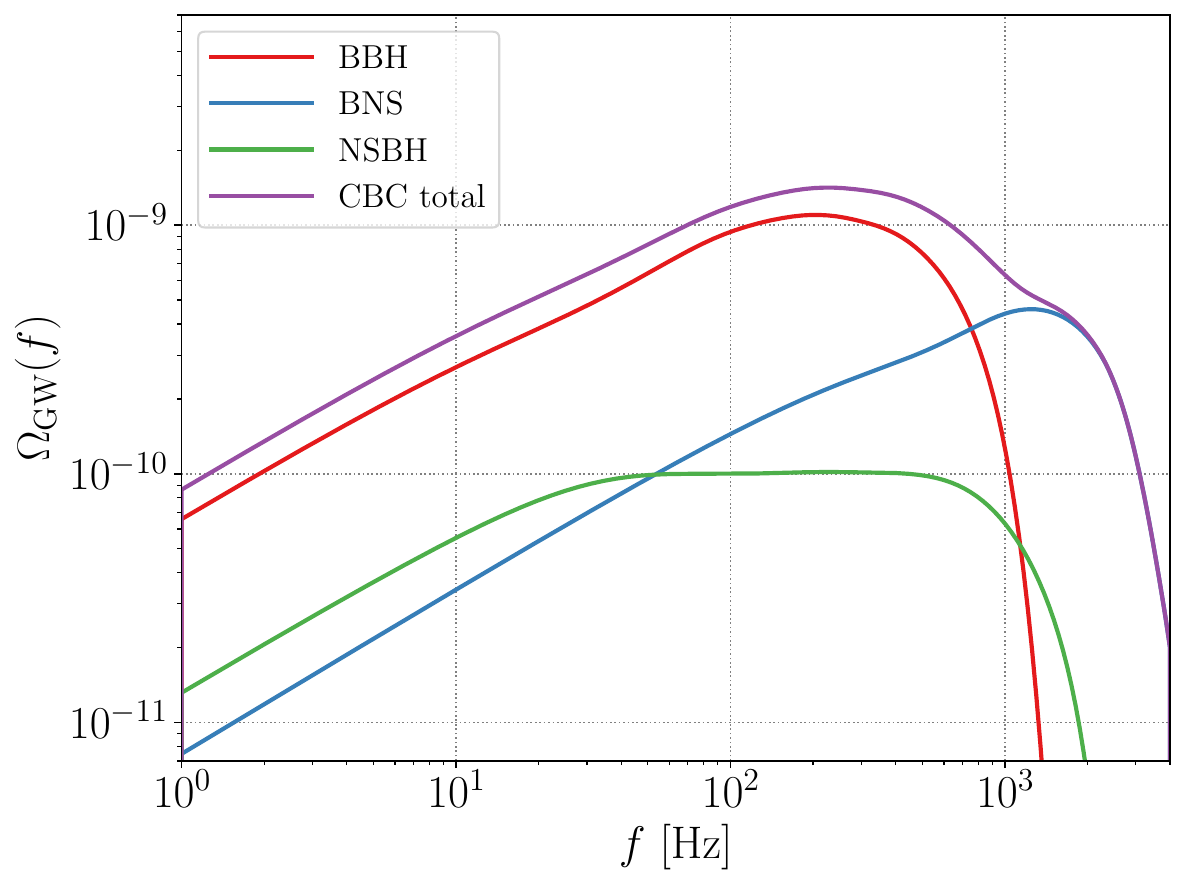}
    \caption{Fiducial GWB model contributions from BBH, BNS and NSBH mergers, along with their combined CBC background.}
    \label{fig:omegafiducial}
\end{figure}

We assume a flat $\mathrm{\Lambda C D M}$ cosmological model and set the following parameters according to the Planck 2018 results~\cite{Planck:2018}: $H_0 = 67.4 \,\mathrm{km} \,\mathrm{Mpc}^{-1} \mathrm{s}^{-1}$, $\Omega_M = 0.315$ and $\Omega_\Lambda = 1 - \Omega_M$. 
In~\cref{fig:omegafiducial}, we present the fractional energy density spectrum of the GWB from the three classes of CBCs, along with their combined contribution, based on the fiducial models.
The dependence of $\Omega_\gw (f)$ on cosmological parameters arises primarily through the $1/H_0^2$ scaling. Additionally, the SFR history is affected via changes in the comoving volume element, while variations in $\Omega_M$ have a negligible effect. 
If, instead, we were to adopt a cosmology based on the SH0ES measurement of the Hubble constant~\cite{Riess:2021}, with $H_0 = 73.04 \,\mathrm{km}\, \mathrm{s}^{-1} \mathrm{Mpc}^{-1}$ and $\Omega_M = 0.262$, this would result in a reduction of the amplitude of $\Omega_\gw$ by approximately 20\%.

\section{Investigating the uncertainty in the binary population}
\label{sec:uncertainty-population}

We investigate the uncertainty in the GWB energy spectrum arising from current constraints on compact binary mass and spin distributions. For BNS systems, we additionally examine the impact of different equations of states (EOSs) on the resulting energy spectrum.

\subsection{BBH mass models} \label{sec:masses}

\begin{figure}[t]
    \centering
    \includegraphics[width=0.48\textwidth]{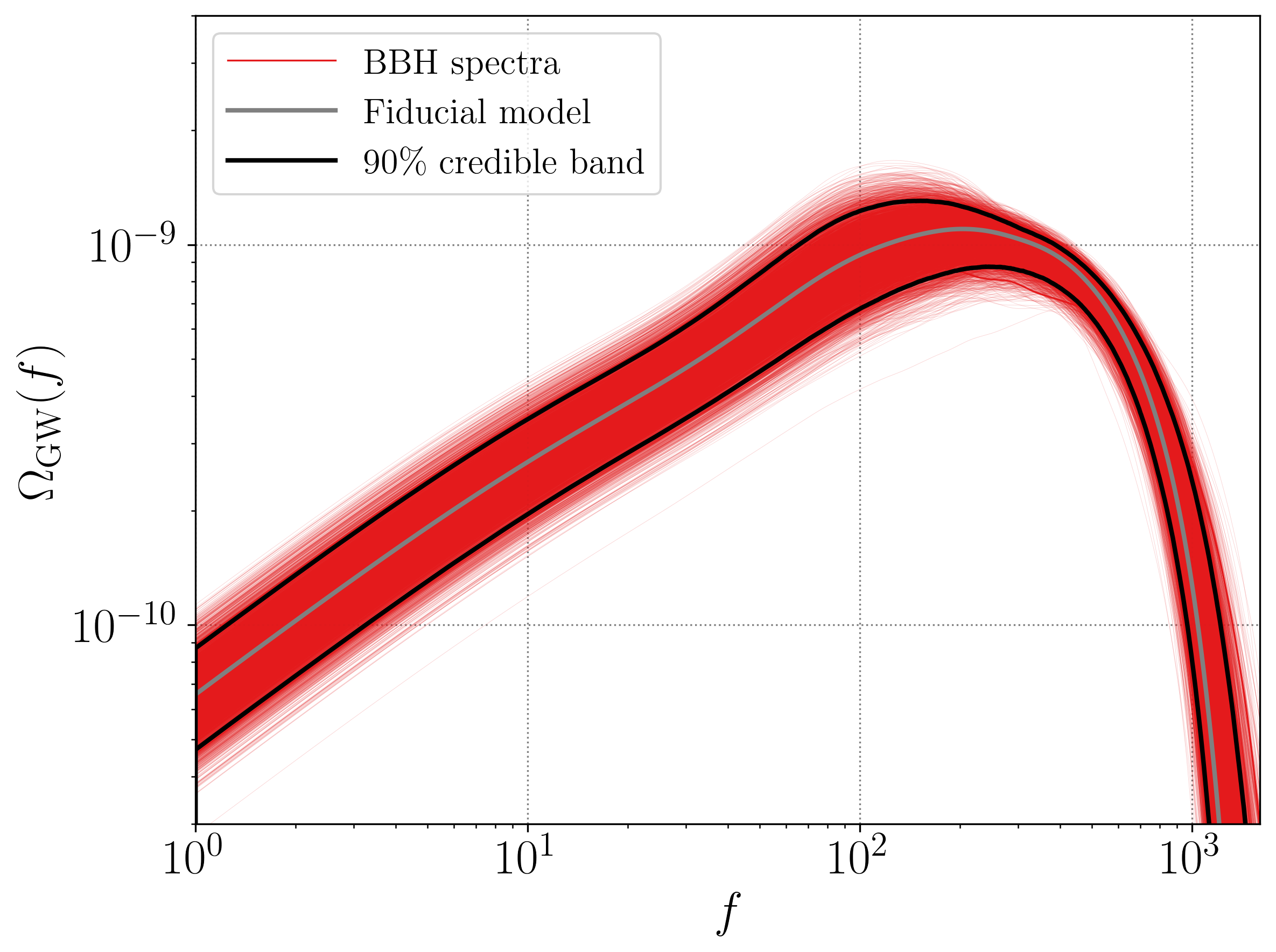}
    \caption{We show the effect of the uncertainty in the BBH mass model on the energy spectrum. The individual spectra correspond to the GWB according to a single hyperparameter posterior sample. The solid black lines show the combined 90\% credible level and the gray line the fiducial BBH background model.}
    \label{fig:PLPmass}
\end{figure}

In~\cref{sec:binarypop}, we introduced the PLP mass distribution model, which provides the best fit to the observed BBH population in GWTC-3~\cite{GWTC3_population}. For our fiducial model we adopted a single realization of this model, using the median values of the hyperparameters inferred from the population analysis. To accurately capture the uncertainties and correlations among these hyperparameters, we incorporate the full set of publicly available posterior samples obtained from the hierarchical Bayesian analysis of BBH mergers in GWTC-3. For each sample, we compute the corresponding GWB energy spectrum. The resulting ensemble of spectra, shown in~\cref{fig:PLPmass}, reflects the full uncertainty associated with the mass distribution. Additionally, we highlight the 90\% credible band arising solely from the uncertainty in the mass distribution.
At lower frequencies, the background amplitude is proportional to the average chirp mass of the BBH population raised to the power of 5/3, where the chirp mass is the combination of component masses $\mathcal{M} = (m_1 m_2)^{3/5} / (m_1+m_2)^{1/5} $. For the PLP mass distribution model, this average is primarily a function of the spectral index $a$ of the power law and $\lambda_p$, the fraction of binaries in the Gaussian peak. A steeper power law leads to more low-mass binaries, while a smaller $\lambda_p$ reduces the contribution from the Gaussian peak near $34~\Msun$. The exact location and width of the Gaussian peak have only a marginal effect.
These two parameters also largely control the shape of the background spectrum near its maximum. While a steeper power law shifts the peak toward higher frequencies, a dominant Gaussian peak tends to broaden it toward lower frequencies.

\begin{figure}[t]
    \centering
    \includegraphics[width=0.48\textwidth]{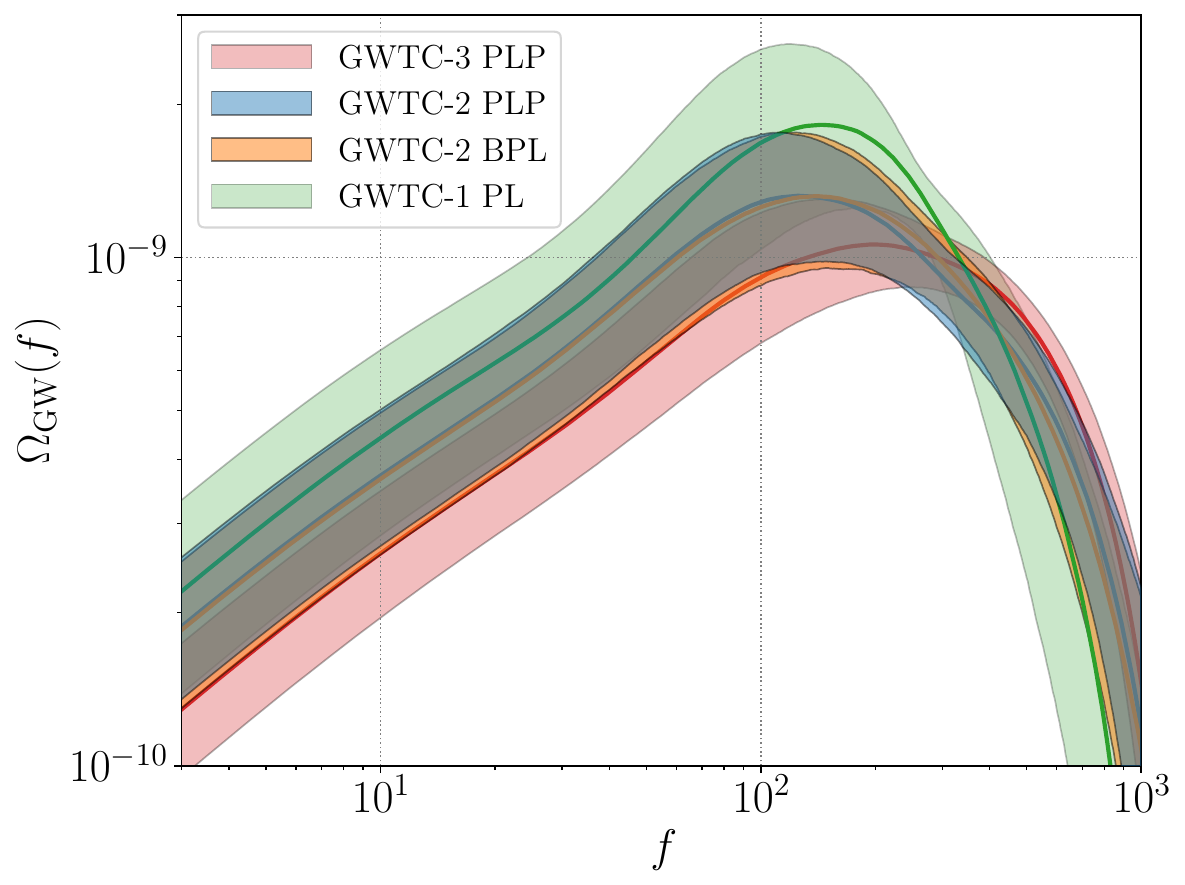}
    \caption{Comparison of the 90\% credible bands of the BBH background spectrum for different mass distribution models. The solid line represents the median within each band. Note that the two GWTC-2 mass models (in blue and orange) show nearly complete overlap.}
    \label{fig:BBHmass}
\end{figure}

To contextualize the uncertainty in the background spectrum derived from the GWTC-3 PLP model, we compare its 90\% credible band with those obtained using different population models and underlying GW data. Specifically, we consider the credible bands computed using the PLP and broken power-law (BPL) models from GWTC-2~\cite{LIGO_GWTC-2:2020,LIGO_O3apop:2020}, as well as the simple power-law model adopted in GWTC-1~\cite{LIGO_GWTC-1:2019,LIGO_O2pop:2019}. This comparison, shown in~\cref{fig:BBHmass}, illustrates how our understanding of the BBH mass distribution has evolved with the increasing number of GW detections. The broader credible intervals in earlier models reflect the larger uncertainty due to fewer BBH observations.
Notably, the two parametric models from GWTC-2, PLP and BPL, result in very similar credible bands.

The mass of merging compact objects does not depend (or depends very mildly) on the merger redshift. Even the mass distribution of black holes depends only mildly on redshift, because BBHs originating from metal-poor progenitors dominate the entire population of merging BBHs across cosmic time~\cite{Mapelli:2019}.

\subsection{BBH spin models} \label{sec:spins}

We investigate the impact of different spin models on the BBH background energy spectrum and compare them to the nonspinning fiducial models. The GWTC-3 population analysis considered two spin models~\cite{GWTC3_population}. The first, referred to as the default model, treats spin magnitudes and orientations independently. Spin magnitudes for both black holes are drawn from a Beta distribution, while tilt angles, defined as the angle between each black hole's spin and the orbital angular momentum, are drawn from a mixture of an isotropic distribution and a preferentially aligned component, modeled by a truncated Gaussian distribution~\cite{Wysocki:2018}.
The second, known as the Gaussian spin model, describes the joint distribution of the effective inspiral spin parameter $\chi_\mathrm{eff}$ and the effective precession spin parameter $\chi_\mathrm{p}$ using a bivariate Gaussian with nonzero correlation~\cite{Miller:2020}.
Using posterior samples of the spin model hyperparameters from the population inference~\cite{GWTC3_population}, we compute the GWB energy spectrum for each sample, modeling spins according to the distributions while keeping all other parameters fixed to those in the fiducial model.
To account for spin precession, we employ the waveform model \texttt{IMRPhenomXP}~\cite{Pratten:2020b}. 

\begin{figure}[t]
    \centering
    \includegraphics[width=0.48\textwidth]{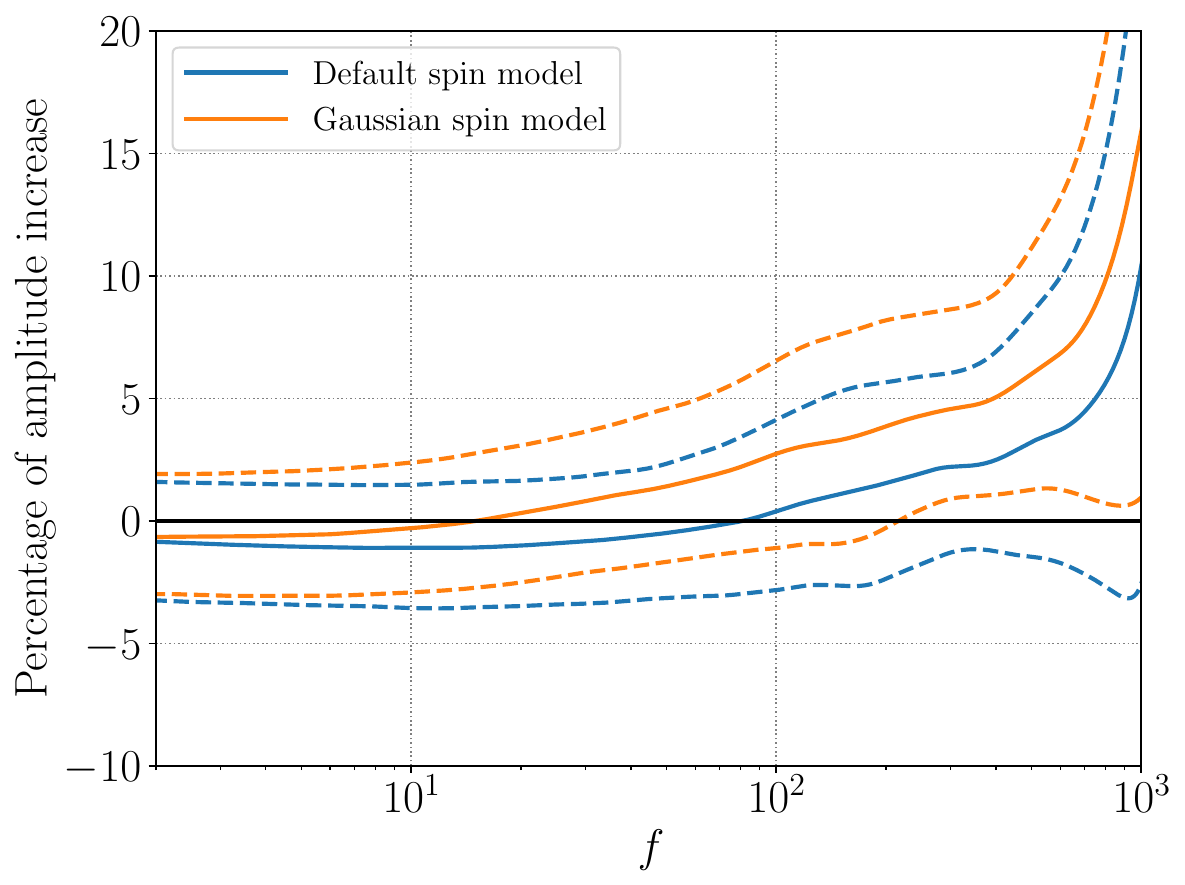}
    \caption{Relative change in the amplitude of the BBH GWB, $\Omega_\mathrm{BBH} (f)$, due to the inclusion of spin effects. The curves show deviations from the nonspinning fiducial model at the 90\% credible level for two spin distribution models, the default spin model and the Gaussian spin model.}
    \label{fig:spin_comp}
\end{figure}

The influence of spin distributions on the stochastic background energy spectrum is notably smaller than that of mass distribution uncertainties. Therefore, we present only the relative difference with respect to the fiducial model in~\cref{fig:spin_comp}. At low frequencies, both spin models cause amplitude uncertainties of approximately 3\%, while at higher frequencies, the background amplitude generally increases, more prominently for the Gaussian spin model. Around 200 Hz, this enhancement can reach up to 10\%. We also note that at low frequencies the spectrum amplitude tends to decrease when spin effects are included. This behavior arises because spins are preferentially aligned~\cite{GWTC3_population}, which accelerates the inspiral and reduces the energy radiated at lower frequencies~\cite{Buonanno:2012}. In contrast, aligned-spin binaries experience the orbital hang-up effect~\cite{Campanelli:2006,Buonanno:2005}, allowing tighter orbits just before merger and emit more GW energy at higher frequencies compared to nonspinning systems.

\subsection{Impact of higher-order modes}

\begin{figure}[t]
    \centering
    \includegraphics[width=0.48\textwidth]{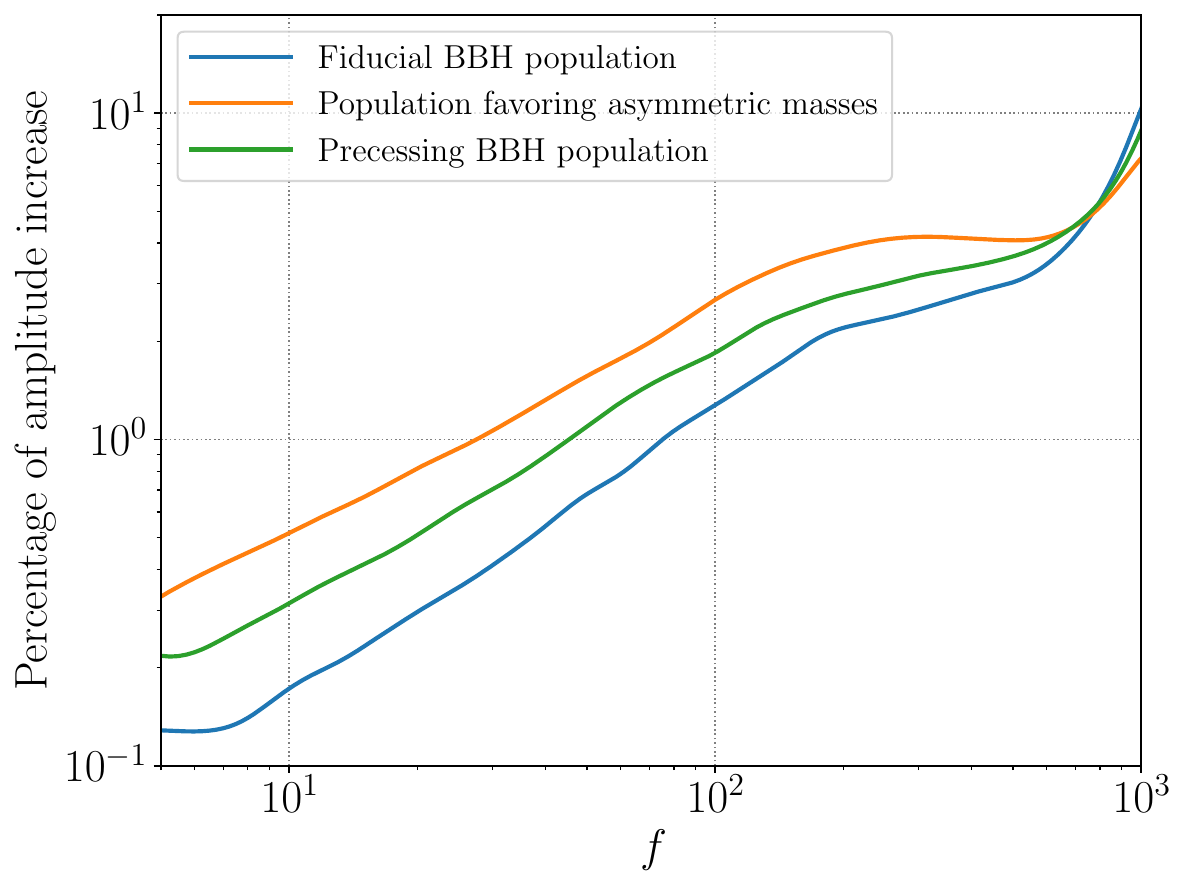}
    \caption{Relative increase in the amplitude of the BBH GWB, $\Omega_\mathrm{BBH} (f)$, when higher-order modes are included using the waveform model \texttt{IMRPhenomXPHM}.}
    \label{fig:omega_hm}
\end{figure}

So far we have computed the background energy spectra using only the dominant (2,2) mode of the waveform. Here, we assess the impact of including higher-order modes by employing the waveform model \texttt{IMRPhenomXPHM}. Since higher-order modes depend differently on the binary's inclination angle than the (2,2) mode, it is necessary to explicitly sample over inclination angles when computing $\Omega_\mathrm{GW}(f)$ via~\cref{eq:OmegaGW}, rather than relying on an isotropic average. 
We first evaluate the fiducial model using \texttt{IMRPhenomXPHM}. As shown in~\cref{fig:omega_hm}, the inclusion of higher-order modes leads to a modest increase in the background amplitude, particularly at higher frequencies near the peak and falloff of $\Omega_\mathrm{GW}(f)$. At lower frequencies, where most BBHs are in the inspiral phase, the effect is negligible.
Next, we keep the PLP mass distribution but select a set of hyperparameters that favor asymmetric masses, specifically, we choose one with a low value of $\beta_q$. Since higher-order modes become more prominent in binaries with unequal masses~\cite{Mills:2020,LIGO_GW190412:2020}, we expect a stronger impact. Comparing the resulting amplitude increase to the same mass distribution evaluated without higher order-modes, we find that the increase is greater than in the fiducial model, though still relatively small.
Furthermore, we consider a model that retains the fiducial mass distribution but incorporates a spin model favoring large-spin magnitudes and tilt angles near $90^\circ$. Such a configuration results in a population of strongly precessing binaries, where the GW signal is modulated and subject to mode mixing, thereby increasing the relevance of higher-order modes~\cite{Pekowsky:2012,Ramos-Buades:2020}. Comparing the resulting spectrum to the one computed with only the dominant mode, we find a marginal amplitude increase as compared to the fiducial case. This outcome is expected: while precession modulates the waveform, it has a limited effect on the total radiated GW energy.

\subsection{BNS mass models}

\begin{figure}[t]
    \centering
    \includegraphics[width=0.48\textwidth]{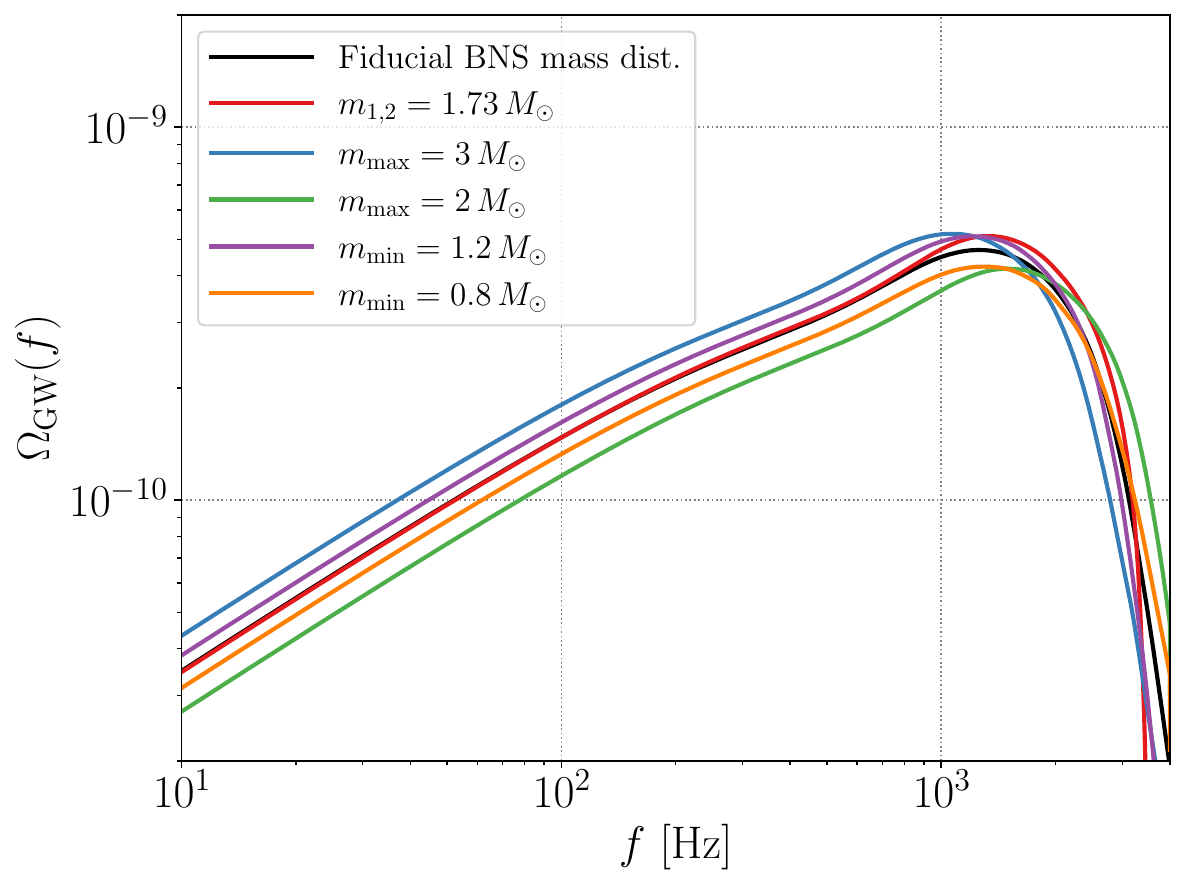}
    \caption{GWB energy density spectra for different BNS mass distributions. The black and red curves compare the fiducial uniform mass distribution with a delta-function distribution centered at the same average chirp mass, showing differences only near the peak. Additional curves illustrate how modifying the minimum or maximum mass of the fiducial distribution affects the shape and amplitude of the spectrum.}
    \label{fig:BNSchirp}
\end{figure}

Over a broad range of the relevant frequency band, BNS systems remain in the inspiral phase, during which the stochastic GWB follows the characteristic spectral dependence $\Omega(f) \sim f^{2/3}$. The amplitude of this background is directly influenced by the population-averaged chirp mass.
Since GW energy emitted by a compact binary scales approximately as $\mathcal{M}^{5/3}$, a higher average chirp mass results in a stronger background signal, particularly at frequencies below the merger regime. Moreover, the distribution of chirp masses affects the spectral shape near the peak of $\Omega(f)$. Lower chirp masses shift the peak to higher frequencies, while higher masses cause deviations from the $f^{2/3}$ scaling at lower frequencies. A broader chirp mass distribution tends to flatten the peak, whereas a narrow distribution produces a more sharply defined peak.
This behavior is illustrated in~\cref{fig:BNSchirp}, where we compare the energy density spectrum for the fiducial uniform mass distribution between 1 and 2.5~$\Msun$ with that of a delta function at the corresponding average chirp mass of 1.73~$\Msun$. 
The two models yield nearly identical amplitudes across most of the spectrum, with noticeable differences only near the peak. 
Additionally, we display how varying the minimum or maximum mass of the fiducial distribution alters the shape and amplitude of the resulting energy density spectrum. 

\begin{figure}[t]
    \centering
    \includegraphics[width=0.48\textwidth]{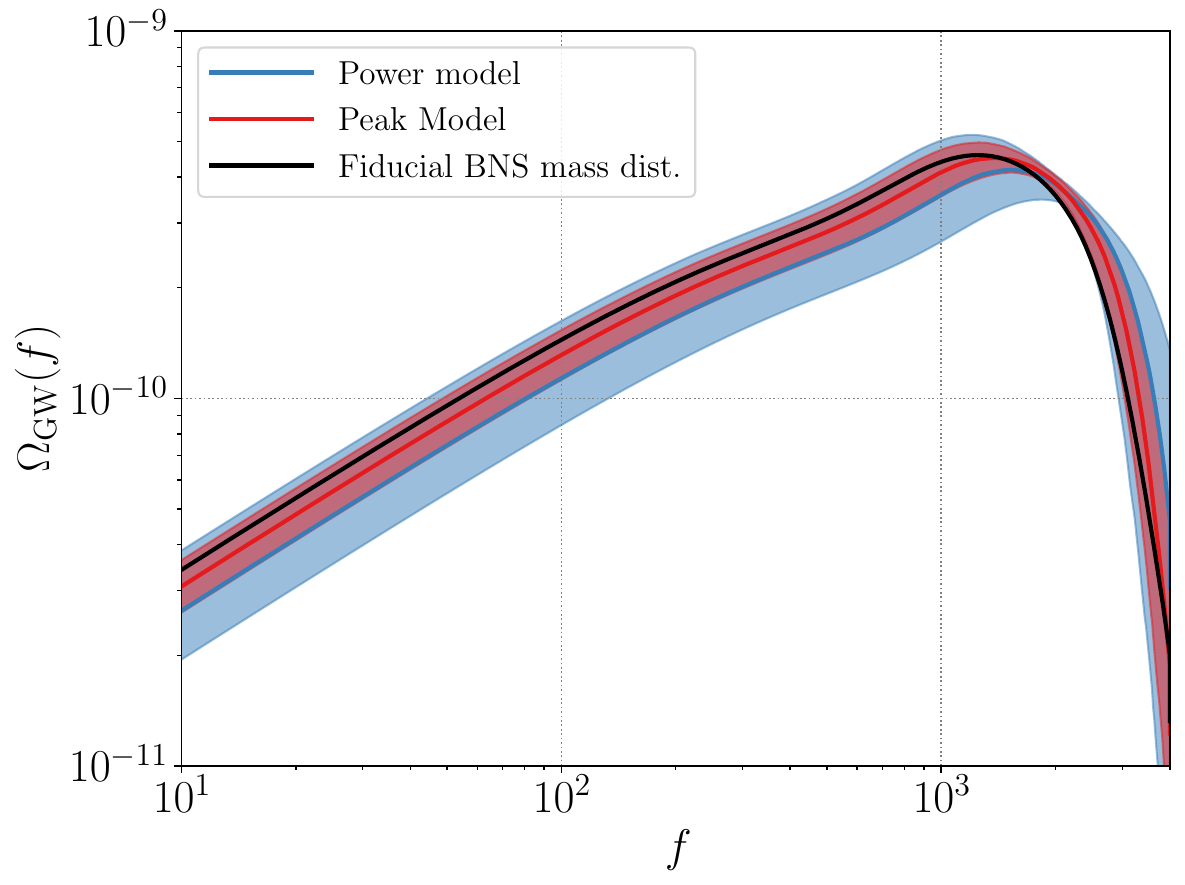}
    \caption{Uncertainty in the GW energy density spectrum from BNS coalescences, based on two different mass distribution models: a power-law model and a Gaussian peak model.}
    \label{fig:BNSmass}
\end{figure}

We investigate the impact of two BNS mass distribution models, presented in Ref.~\cite{GWTC3_population}, on the GW energy density spectrum and its associated uncertainty. The first model assumes a power-law distribution for the component masses $p(m) \propto m^\alpha$ within the range $m_\mathrm{min} < m < m_\mathrm{max}$. The 90\% credible intervals for the inferred parameters are: $\alpha = -2.1^{+5.2}_{-6.9}$, $m_\mathrm{min} = 1.2_{-0.2}^{+0.1} \,\Msun$, and $m_\mathrm{max} = 2.0^{+0.3}_{-0.3} \,\Msun$, though we choose $2.5 \,\Msun$ for the upper mass limit to explicitly include the fiducial BNS mass distribution. This fiducial model corresponds to a uniform distribution (i.e., $\alpha = 0$) between $1\, \Msun$ and $2.5\, \Msun$. 
In the second mass model, the component masses are drawn from a truncated Gaussian with mean $\mu$ and standard deviation $\sigma$ between some minimum and maximum mass. The inferred parameters are: $\mu = 1.5^{+0.4}_{-0.3} \,\Msun$, $\sigma = 1.1^{+0.8}_{-0.8}\, \Msun $, $m_\mathrm{min} = 1.1_{-0.1}^{+0.2} \,\Msun$, and $m_\mathrm{max} = 2.0^{+0.2}_{-0.2} \,\Msun$. As with the power-law model, we set the upper mass limit to $2.5\, \Msun$ to match the fiducial range.
In~\cref{fig:BNSmass} we show the 90\% credible uncertainty in the energy density spectrum introduced by these mass distribution models. The power-law model allows for a broader variation in the spectrum compared to the Gaussian (or "Peak") model.

\subsection{BNS equation of state}

\begin{figure}[t]
    \centering
    \includegraphics[width=0.48\textwidth]{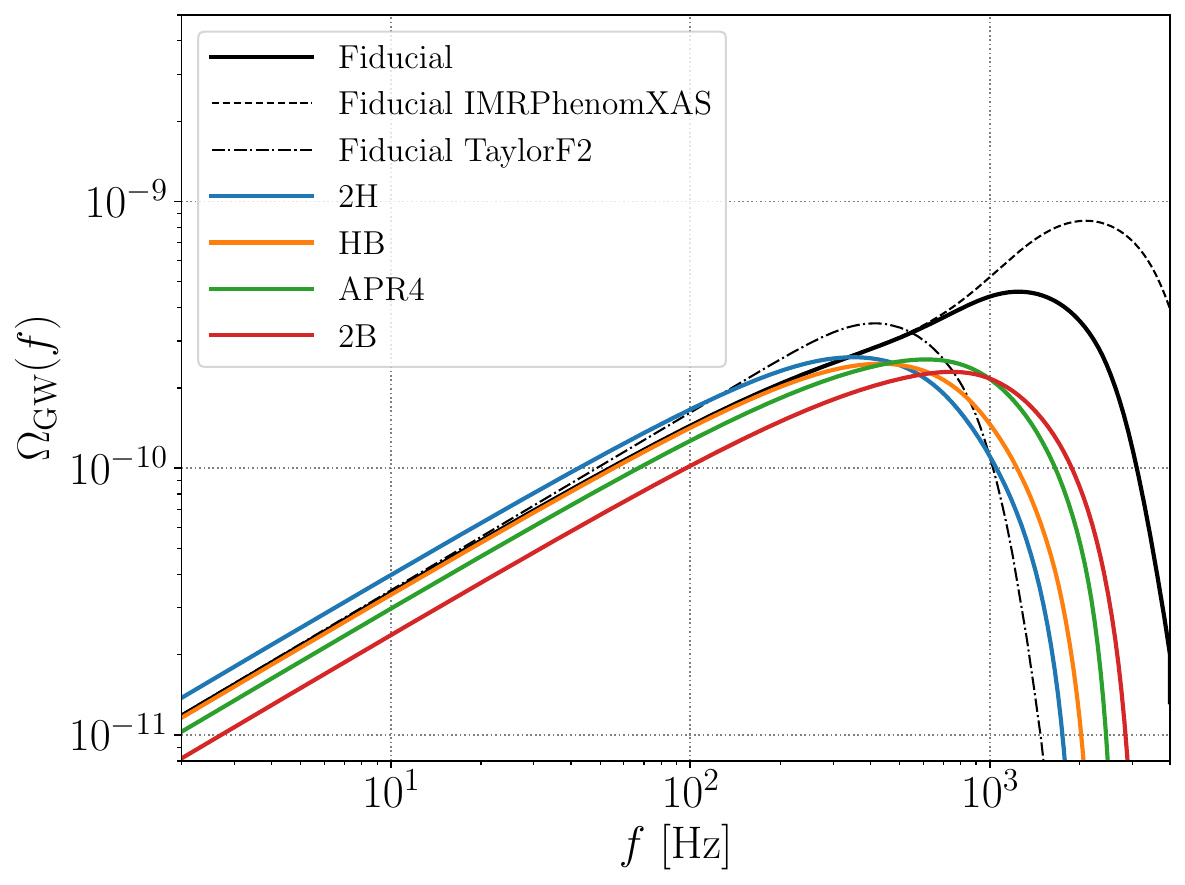}
    \caption{GWB energy density spectrum $\Omega_\gw (f)$, for different neutron star EOSs, assuming identical parameters to the fiducial BNS model. The fiducial model is shown alongside evaluations using a BBH waveform and an inspiral-only waveform.}
    \label{fig:BNSeos}
\end{figure}

The EOS describes the relationship between pressure and density in neutron star matter, thereby determining the mass-radius relation and other macroscopic properties, such as tidal deformability. As its precise form is still largely unknown, we investigate how different EOS models influence the energy density spectrum.
As in the fiducial model, we assume a uniform distribution of component masses in the range from $1 \, \Msun$ to $m_\mathrm{max,TOV}$, the maximum mass of a neutron star supported by the selected EOS.
To explore a broad region of the EOS parameter space, we consider four representative EOS models: 2H~\cite{Read:2008}, HB~\cite{Hebeler:2013}, APR4~\cite{Akmal:1998}, and 2B~\cite{Read:2008}. These span from stiff EOSs, which predict larger neutron star radii and higher maximum masses, to softer EOSs, which yield more compact stars with lower maximum masses. For example, the APR4 EOS supports a maximum mass of $m_\mathrm{max,TOV} = 2.2\, \Msun$ and its tidal deformability at the reference mass of $1.4 \, \Msun$ is $\Lambda_{1.4} = 257$. 
In~\cref{fig:BNSeos}, we present the resulting energy density spectra for the different EOSs. Two distinct effects can be observed: first, an increase in amplitude at lower frequencies, which is directly related to the maximum mass supported, and second, a shift of the spectral peak to lower frequencies for stiffer EOSs. This effect is mainly due to the fact that larger neutron stars come into contact earlier during the inspiral, leading to a tapering of the waveform.
Additionally,~\cref{fig:BNSeos} illustrates the impact of different waveform models on $\Omega_\gw(f)$. Compared to the BBH waveform model \texttt{IMRPhenomXAS}, the BNS model exhibits tapering in the merger and ringdown phases, even when no tidal deformability is assumed. In contrast, the inspiral-only approximant \texttt{TaylorF2}~\cite{Messina:2019}, which terminates at the last stable circular orbit, deviates from the fiducial model at higher frequencies by omitting contributions from the merger and ringdown.
Finally, we note that an additional EOS-related effect, not considered here, is the potential contribution of a postmerger signal, which could impact the background in the kHz regime~\cite{Lehoucq:2025}.

\subsection{NSBH mass models}

\begin{figure}[t]
    \centering
    \includegraphics[width=0.48\textwidth]{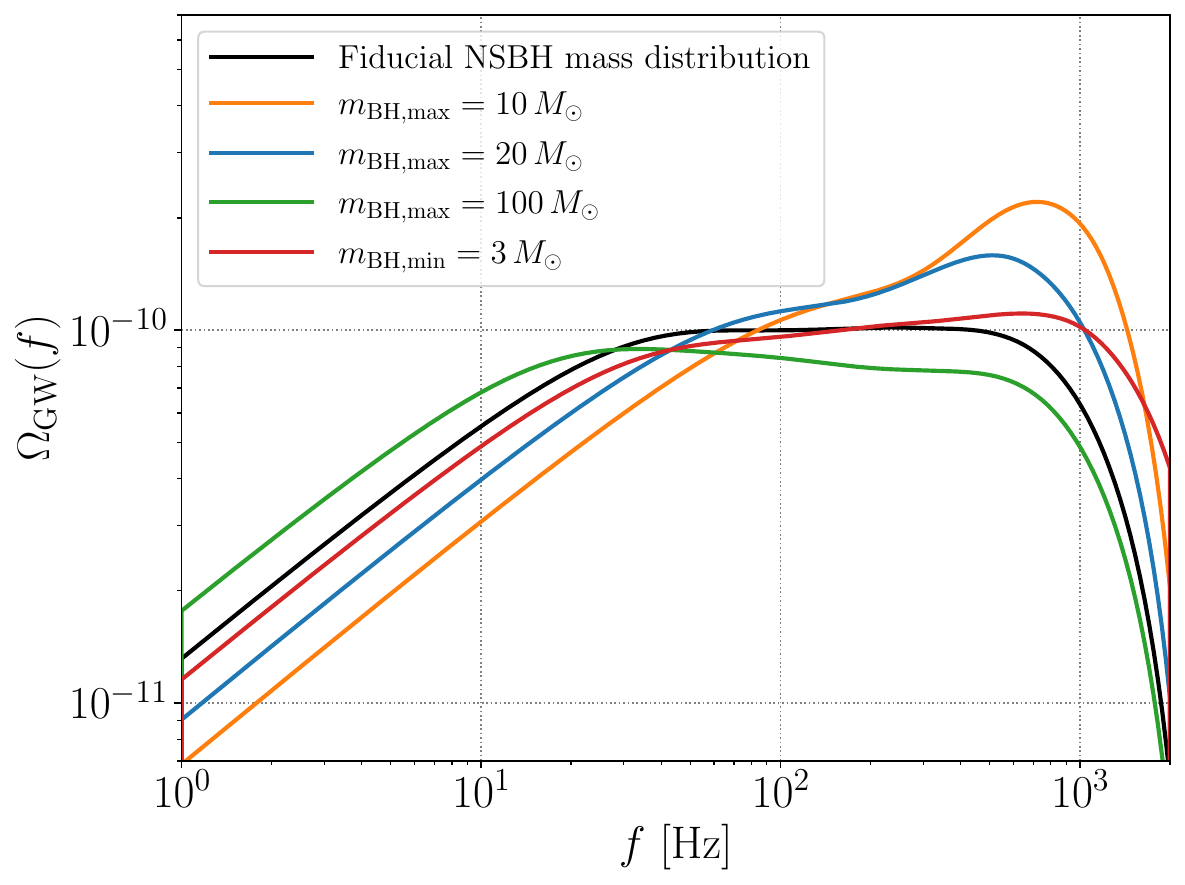}
    \caption{GWB energy density spectrum $\Omega_\gw (f)$ for NSBH mergers with varying black hole mass distributions. The black hole masses follow log-uniform distributions with different upper and lower limits, while the neutron star mass is uniformly distributed between $1.0\, \Msun$ and $2.5\, \Msun$. Legend entries indicate only the deviation from the fiducial model, which assumes black hole mass limits of $5\, \Msun$ and $50\, \Msun$.}
    \label{fig:NSBHmass}
\end{figure}

The spectral shape of the GWB from NSBH mergers is highly sensitive to the assumed upper limit of the black hole mass when the mass distribution is assumed to be log-uniform, as illustrated in~\cref{fig:NSBHmass}. At low frequencies, we observe the characteristic $f^{2/3}$ scaling of inspiral-dominated signals, with its amplitude governed by the population-averaged chirp mass, as discussed previously. 
For instance, setting the upper mass limit to $100 \,\Msun$ leads to a flattening of the spectrum already at around 10~Hz, and it can even peak around 30~Hz due to the increased contribution from high-mass systems merging at lower frequencies. 
In contrast, when the maximum mass is restricted to $10\,\Msun$, the inspiral regime extends to higher frequencies, resulting in a steeper spectrum and a substantially enhanced amplitude toward 1~kHz. 
The influence of the neutron star mass distribution on the NSBH energy density spectrum is comparatively minor.
An important caveat not addressed here is the role of tidal interactions, which can significantly affect the GW energy emitted~\cite{Flanagan:2007,Vines:2010}. 
In particular, tidal disruption of the neutron star may occur when the masses of the black hole and the neutron star are not too unequal, a condition that depends sensitively on the neutron star's EOS~\cite{Kyutoku:2010,Kyutoku:2011,Foucart:2012}. 
In such cases, the neutron star can be disrupted before merger, leading to an earlier termination of the waveform and suppression of high-frequency GW emission~\cite{Pannarale:2015}. As a result, using a BBH waveform model may overestimate the contribution at higher frequencies to the GWB.
Conversely, for large mass ratios, the neutron star typically plunges intact into the black hole, and the waveform is well approximated by a BBH model~\cite{Foucart:2013,Matas:2020}. 

\section{Investigating the uncertainty in the merger rate history}
\label{sec:uncertainty-mergerrates}

Estimating compact binary merger rates requires modeling the complex interplay between star formation and the evolution of binary star systems. This involves tracking the formation and evolution of binaries through processes such as mass transfer, stellar winds, and supernova explosions, which eventually produce compact objects like neutron stars and black holes. The merger rate is then determined by considering the timescales for these binaries to coalesce due to gravitational wave emission, along with factors like metallicity and the initial mass function. 
This approach enables predictions of merger rates across cosmic time, particularly at high redshifts where current GW observatories lack sensitivity for direct detections.
In the following, we investigate some specific models for the merger rate history of compact binaries and assess their impact on the GWB energy density spectrum.

\subsection{Time delay}

We start by investigating the effect of the time-delay distribution on the resulting merger rate. In~\cref{eq:Rmz} we have introduced the calculation of the merger rate from the SFR convolved with a distribution of time delays $p(t_d)$. For the fiducial model we have assumed $p \sim t_d^{-1}$, which is motivated by the gravitational wave inspiral timescale assuming the initial orbital separation of binaries is flat in log-space~\cite{Piran:1992}. 
To reflect the uncertainty in the time-delay distribution, we examine $p(t_d) \sim t_d^{\nu}$ for $\nu \in [-1.5, -0.5]$. 
The other critical parameter is the minimum delay time $t_\mathrm{d,min}$, which for the fiducial model was assumed to be 50 Myr. In the literature~\cite{Dominik:2012,Dominik:2013,Belczynski:2016,Rodriguez:2016,Marchant:2016,Artale:2019,Zevin:2022}, values for $t_\mathrm{d,min} \in [10, 100]$ Myr are common, depending on the formation channel and assumptions in stellar evolution and population synthesis modeling.

The first column of~\cref{fig:SFRtimemetal} displays how extreme choices for the parameters $\nu$ and $t_\mathrm{d,min}$ affect the fiducial BBH merger rate (without metallicity cut) and the corresponding background energy density spectrum. A steeper delay-time distribution ($\nu < -1$) favors shorter delay times, leading to a higher peak merger rate and an enhanced background amplitude. Conversely, a flatter distribution or a larger minimal delay time tends to suppress the peak merger rate, which also shifts toward lower redshifts.
The relative difference in background amplitude $\Omega_\mathrm{GW}$ at 10~Hz between the extreme cases ($\nu=-3/2,\,t_\mathrm{d,min}=10\,\mathrm{Myr}$ and $\nu=-1/2,\,t_\mathrm{d,min}=100\,\mathrm{Myr}$) exceeds 130\%, more than doubling the amplitude.
Notably, the shape of the merger rate at low redshifts is particularly sensitive to the delay-time distribution. With upcoming observing runs, we anticipate more precise measurements of the local BBH merger rate and the low-redshift spectral index $\gamma$, which will help constrain the time-delay models.

\subsection{Metallicity-weighted SFR}

In the past, different SFRs have been used to compute the binary merger rate and then the resulting background energy density spectrum. Here, we explicitly show in three examples that the form of the underlying SFR has a small effect compared to the delay-time distribution or metallicity cuts.
We compare the Madau-Fragos SFR~\cite{Madau:2016} introduced in~\cref{eq:SFR} to the Madau-Dickinson SFR~\cite{Madau:2014}, which has the same functional form but slightly different parameter values, $\gamma = 2.7$, $\kappa = 5.6$, and $z_\mathrm{peak} = 1.9$.  
Additionally, we consider the SFR provided in Ref.~\cite{Vangioni:2014} (Vangioni), which is based on the gamma-ray burst rate and is parametrized by the following form given in Ref.~\cite{Springel:2002}:
\begin{align}
    R_f (z) \propto \frac{a \exp (b(z-z_m))}{a-b+b \exp (a(z-z_m))} \,,
\end{align}
which was used, for example, in the computation of the CBC background of Ref.~\cite{LVK-isotropic-o3:2021}. Here, we consider the following values: $z_m =1.72$, $a=2.8$, and $b=2.46$.
In the middle panel of~\cref{fig:SFRtimemetal} we illustrate the impact of different SFR models on the BBH merger rate and the resulting GW energy density spectrum, both with and without the metallicity cut defined in~\cref{eq:gammetal}. While the Vangioni SFR model predicts a higher merger rate at large redshifts, its influence on the energy density spectrum is minimal. Similarly, the minor changes in the parameter values from the Madau-Dickinson to the Madau-Fragos model do not significantly alter the background.

\subsection{Metallicity model}

Here, we investigate a particular metallicity model to compute the merger rate of compact binaries from the metallicity-weighted SFR~\cite{Belczynski:2016,Madau:2014}. We adopt a prescription of the mean metallicity evolution of the Universe based on the assumption that the average metallicity of star-forming gas increases with cosmic time, reflecting the cumulative enrichment by successive generations of stars. The mean metallicity at each redshift is given by
\begin{align} \label{eq:meanmetal}
    &\log \left(Z_{\text {mean }}(z)\right) = 0.5 \nonumber \\ &\qquad+ \log \left(\frac{y(1-R)}{\rho_b} \int_z^{z_\mathrm{max}} d z^{\prime} \,\frac{R_f\left(z^{\prime}\right)}{ H\left(z^{\prime}\right)\left(1+z^{\prime}\right)} \right) \,,
\end{align}
where $R=0.27$ is the return fraction, a net metal yield $y = 0.019$, a baryon density $\rho_b = 2.77 \times 10^{11} \Omega_b h^2 \, \Msun \mathrm{Mpc}^{-3}$ with $\Omega_b h^2 = 0.0224$, and the specific Madau-Fragos SFR $R_f(z)$~\cite{Madau:2016}.
At each redshift we assume a log-normal distribution of metallicity around the mean, with standard deviation $\sigma_Z$. 
Since BBHs form preferentially in low-metallicity environments, we can introduce a threshold metallicity $Z_\mathrm{th}$ beyond which no BBH formation is possible. The fraction of stars $\varepsilon_Z (z)$ formed at redshift $z$ with metallicities below $Z_\mathrm{th}$ can be expressed as
\begin{align}
    \varepsilon_Z (z) = \frac{1}{2} \left[1 + \operatorname{erf} \left( \frac{\log Z_\mathrm{th} - \log (Z_\mathrm{mean} (z))}{\sqrt{2} \, \sigma_Z} \right) \right] \,.
 \end{align}
Here, we set the solar metallicity $Z_\odot = 0.014$ according to Ref.~\cite{Asplund:2009}.
Now we can multiply the SFR with $\varepsilon_Z (z)$ to get the metallicity-weighted SFR, which is heavily influenced by the choice of $Z_\mathrm{th}$ and $\sigma_Z$.

\begin{figure*}[t]
    \centering
    \includegraphics[width=0.98\textwidth]{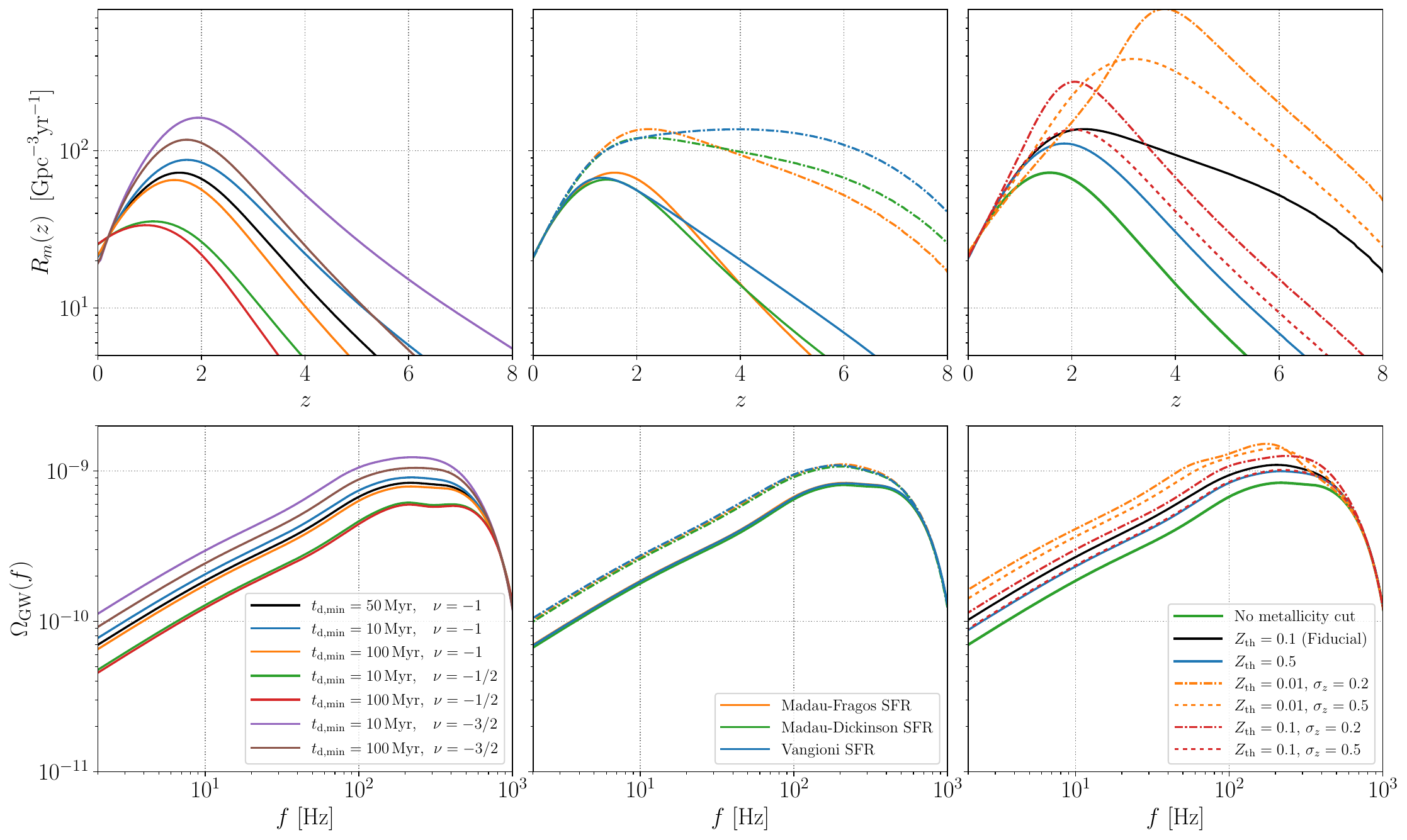}
    \caption{Impact of different modeling choices on the BBH merger rate and the resulting GWB energy density spectrum. 
    \textit{Left panel:} effect of varying the time-delay distribution. Shorter minimum delay times and steeper distributions lead to higher merger rates and enhanced background amplitudes. 
    \textit{Middle panel:} influence of different SFR models. We compare the merger rates derived from the Madau-Fragos~\cite{Madau:2016}, Madau-Dickinson~\cite{Madau:2014}, and Vangioni~\cite{Vangioni:2014} SFR models, shown both with (dashed lines) and without (solid lines) the fiducial metallictiy cut. 
    While the choice of SFR has a minor effect, applying the metallicity cut substantially enhances the merger rate and the resulting GWB amplitude. 
    \textit{Right panel:} effect of metallicity modeling. We compare the fiducial metallicity model using two different thresholds with the metallicity model from Ref.~\cite{Belczynski:2016} evaluated for two thresholds, each with two values of metallicity dispersion. For reference we also show the case without any metallicity cut. 
    Stricter metallicity thresholds and narrower dispersions shift the peak of the merger rate to higher redshifts and increase its amplitude, thereby enhancing the predicted GW background.}
    \label{fig:SFRtimemetal}
\end{figure*}

In the right panel of~\cref{fig:SFRtimemetal}, we illustrate how varying the threshold metallicity $Z_\mathrm{th}$ and the metallicity dispersion $\sigma_Z$ influences the merger rate, computed under the fiducial time-delay model.
We also show the corresponding impact on the GWB energy-density spectrum, comparing results to both the fiducial metallicity model [\cref{eq:gammetal}] and the case where no metallicity cut is applied.
Introducing a metallicity cut generally shifts the peak of the merger rate to higher redshifts, allowing for larger maximum rates. This shift, in turn, enhances the predicted energy density spectrum of the GWB.
Specifically, a low threshold $Z_\mathrm{th}$ combined with a narrow dispersion $\sigma_Z$ around the mean metallicity leads to elevated merger rates at earlier cosmic times.
This occurs because in this case, the local Universe contains relatively little low-metallicity star formation, thus, local mergers predominantly originate from the long time-delay tail of a high-redshift progenitor population. As a result, the GWB spectrum exhibits a higher overall normalization and a modified spectral shape near its peak, due to the redshifting of the dominant high-$z$ contribution to lower frequencies.

\subsection{Combined merger rate uncertainty}

\begin{figure*}[t]
    \centering
    \includegraphics[width=0.98\textwidth]{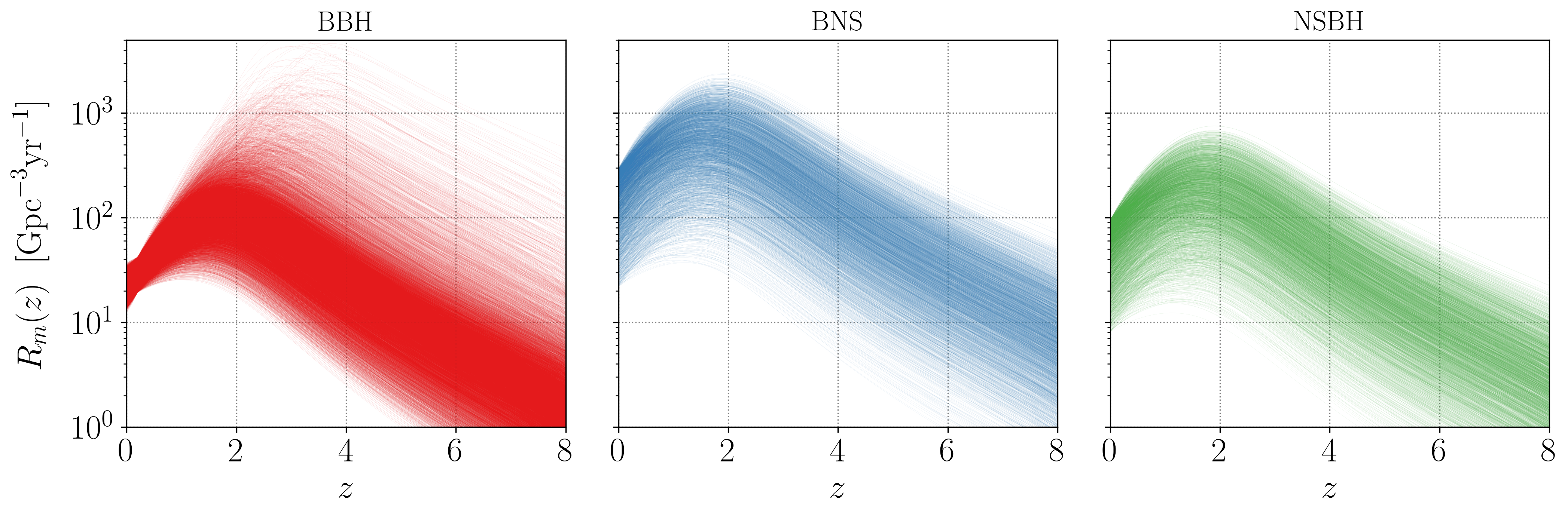}
    \caption{Ensembles of merger rates for BBH, BNS, and NSBH, modeled using the SFR convolved with a time-delay distribution. For BBHs, a metallicity threshold is additionally applied. The parameters describing these models are listed in~\cref{tab:mratevals}. All rates are normalized to local merger rate estimates within the 90\% credible interval from GW observations, following Ref.~\cite{GWTC3_population}.}
    \label{fig:CBCmergerRates}
\end{figure*}

To construct the full ensemble of merger rate histories across all classes of compact binaries, we draw random samples from the parameter space defined earlier, which governs the evolution of the merger rate.
The sampling ranges and underlying distributions for these parameters are summarized in~\cref{tab:mratevals}.
Specifically, we vary the local rate $R_0$, the minimum delay time $t_\mathrm{d,min}$, the spectral index $\nu$ of the time-delay distribution, and exclusively for BBH systems, the metallicity threshold $Z_\mathrm{th}$, above which BBH formation is suppressed, and the metallicity spread $\sigma_Z$, which characterizes the dispersion around the mean metallicity given by~\cref{eq:meanmetal}. 
We adopt a Gaussian prior for $\nu$ (with $\mu_\nu = -1$ and $\sigma_\nu = 0.25$) to reflect results from observations that a spectral index $\nu=-1$ provides a good fit to many compact binary formation models~\cite{McCarthy:2020,Fishbach:2021,Turbang:2023}. This choice helps to avoid a bias toward high merger rates that could arise by large negative values of $\nu$, without affecting the overall uncertainty in the combined merger rate uncertainty. It simply ensures a greater representation of merger rates consistent with a time-delay distribution with spectral index close to $-1$.
While BNS formation is only weakly sensitive to metallicity, the situation is less clear for NSBH systems. Nevertheless, given the already broad uncertainties in the NSBH merger rate estimates, we choose not to introduce additional uncertainty from metallicity effects.
The resulting CBC merger rates are shown in~\cref{fig:CBCmergerRates}. 
While the local BBH merger rate is relatively well constrained by direct observations, the rate at redshifts $z > 2$ can vary by more than two orders of magnitude depending on the model assumptions. The local rates of BNS and NSBH coalescences are much less constrained.

\begin{table}[H]
    \centering
    \setlength{\tabcolsep}{7pt} % Horizontal padding (default is 6pt)
    \renewcommand{\arraystretch}{1.4} % Vertical padding (default is 1)
    \begin{tabular}{|c|c|c|}
    \hline
        $R^\mathrm{BBH}_0 \; [\mathrm{Gpc}^{-3}\,\mathrm{yr}^{-1}$]  & [19.2, 42.2] & uniform \\
        $R^\mathrm{BNS}_0 \; [\mathrm{Gpc}^{-3}\,\mathrm{yr}^{-1}$]  & [22, 296] & uniform \\
        $R^\mathrm{NSBH}_0 \; [\mathrm{Gpc}^{-3}\,\mathrm{yr}^{-1}$]  & [8, 96] & uniform \\
        $t_\mathrm{d,min} \; [\mathrm{Myr}]$ & [0.01, 0.1] & uniform \\
        $\nu $ & [-1.5, -0.5] & truncated Gaussian\\
        $Z_\mathrm{th} \; [Z_\odot]$ & $[0.01, 1.0]$ & uniform \\
        $\sigma_Z \; $ & $[0.2, 0.5]$ & uniform \\
        \hline
    \end{tabular}
    \caption{Sampling intervals and underlying distributions for the parameters used to construct the merger rate histories. The metallicity-related parameters $Z_\mathrm{th}$ and $\sigma_Z$ are only relevant for the BBH case.}
    \label{tab:mratevals}
\end{table}

\section{Combining uncertainties in modeling the astrophysical GWB}
\label{sec:combined-uncertainty}

Having investigated the impact of different models and uncertainty individually, we now combine everything together to get a better estimate of the current uncertainty in the stochastic background energy spectra from different classes of CBCs.

%\subsection{BBH background}

To quantify the current uncertainty in the BBH background spectrum, we construct a credible band based on the ensemble of merger rate histories shown in the left panel of~\cref{fig:CBCmergerRates}. 
Each merger rate history is combined with a set of hyperparameters describing the best-fit models for the BBH mass and spin distributions. Specifically, we adopt the PLP model for the primary mass, a power-law model for the secondary mass, and the default spin model. The hyperparameter posterior samples are obtained via hierarchical Bayesian inference, as detailed in Ref.~\cite{GWTC3_population}.
For each sample, we compute the corresponding BBH background spectrum by evaluating~\cref{eq:OmegaGW}. 
This involves summing contributions from a large number of BBH mergers, with masses, spins, and redshifts drawn according to the population models described by the set of hyperparameters. 
This approach captures the combined uncertainty arising from both the merger rate evolution and the population properties of BBH systems.

For BNS and NSBH systems, we estimate the uncertainty in the background in a similar manner. We combine the rates, which are presented in the middle and right panels of~\cref{fig:CBCmergerRates}, with hyperparameters describing the mass distributions. Since these distributions remain largely unconstrained for these classes of CBCs, we consider the same simple models used for the fiducial case, with some more flexible boundaries. 
For BNS systems, we assume a uniform mass distribution between $m_\mathrm{min}$ and $m_\mathrm{max}$ for both components, where $m_\mathrm{min}$ and $m_\mathrm{max}$ are independently drawn from uniform distributions over the intervals $ [1.0,1.3] \,\Msun$ and $ [2.0,2.5] \,\Msun$, respectively. 
For NSBH systems, the neutron star mass distribution is taken to be the same as in the BNS case, while the black hole mass follows a log-uniform mass distribution with lower and upper bounds drawn uniformly from $ [3,5] \,\Msun$ and $ [10,100] \,\Msun$, respectively.
Given that spin contributions are expected to have minimal impact relative to the existing uncertainties, we assume both BNS and NSBH systems to be nonspinning.

\begin{figure*}[ht!]
    \centering
    \includegraphics[width=0.98\textwidth]{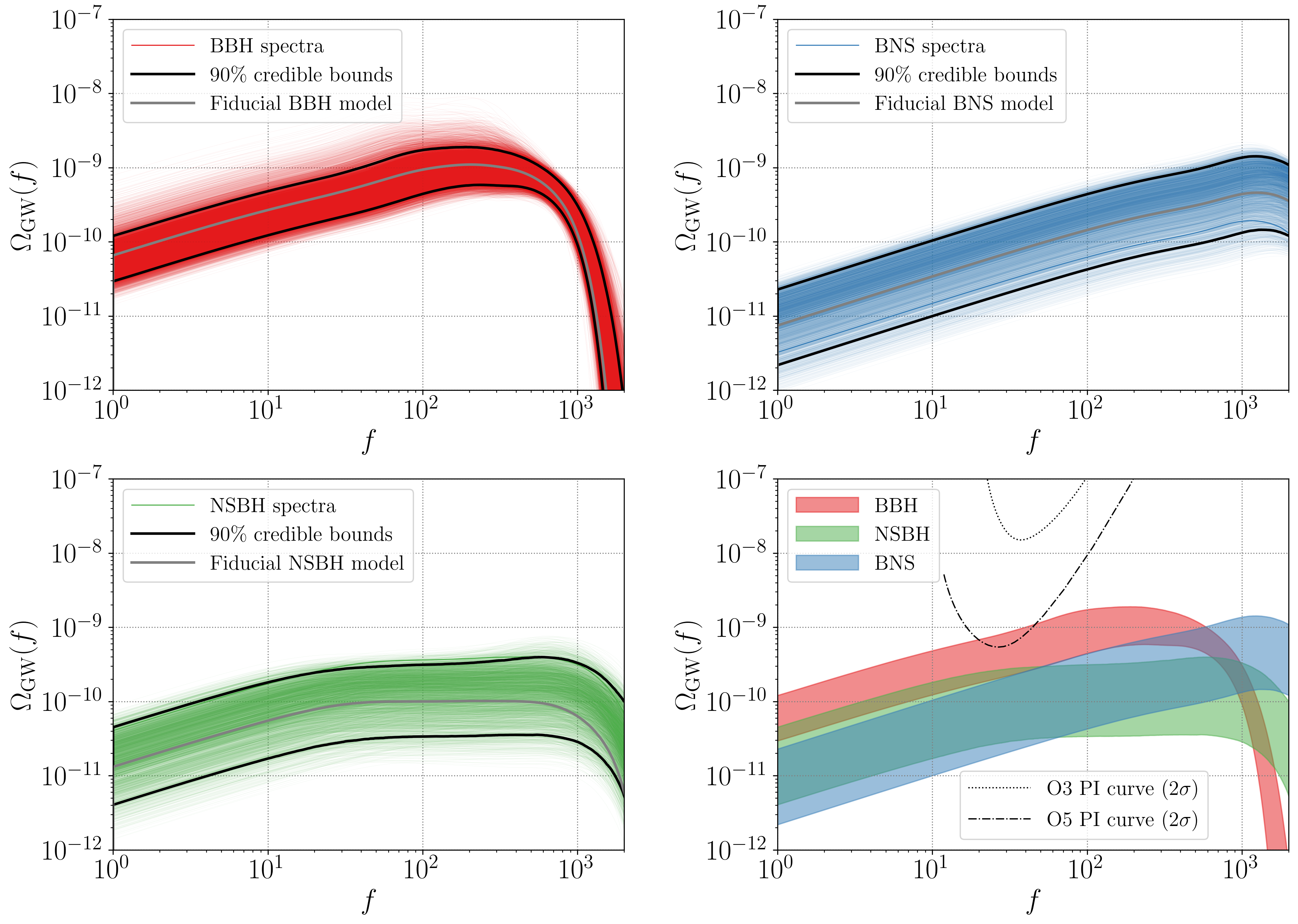}
    \caption{\textit{Upper left:} BBH background spectra derived from a wide ensemble of redshift evolution models, combined with uncertainties in the mass and spin distributions, based on the PLP mass model and the default spin model. Black lines indicate the 90\% credible band, providing a robust estimate of the stochastic background due to BBH mergers. 
    \textit{Upper Right:} BNS background spectra, incorporating merger rate uncertainties (as shown in~\cref{fig:CBCmergerRates}), and a uniform mass distribution with slightly varying bounds. Black lines indicate the 90\% credible band, and the fiducial BNS model is shown in gray. 
    \textit{Lower Left:} NSBH background spectra, combining merger rate uncertainties with slight variations of the fiducial NSBH mass distribution. In gray we indicate the fiducial NSBH background, and black lines show the 90\% credible bounds.
    \textit{Lower Right:} 90\% credible bands for BBH, BNS, and NSBH backgrounds. Also shown are the PI sensitivity curve for the search with data up to O3, and the target PI sensitivity curve anticipated for O5 at the $2\sigma$ level.}
    \label{fig:CBCfull}
\end{figure*}

The resulting spectra are presented in~\cref{fig:CBCfull}, which illustrates both the individual realizations and the 90\% credible band for the three classes of CBCs, alongside the fiducial models introduced earlier. In the bottom right panel of~\cref{fig:CBCfull}, we show again the 90\% credible bands of all CBC backgrounds, together with the $2\sigma$ power-law integrated (PI) curves~\cite{Thrane:2013} from the stochastic cross-correlation search following the third observing run~\cite{LVK-isotropic-o3:2021}, as well as the projected sensitivity for one year of coincident observation at O5 sensitivity~\cite{O5-ligo-psd,O5-virgo-psd} with upgraded Advanced LIGO detectors (A+)~\cite{aplus_design} and the Virgo detector (V+)~\cite{VIRGO:2023}. 
The PI curve indicates the sensitivity to a power-law signal integrated over the entire frequency band at a given confidence level (here, $2\sigma$). If a signal lies above the PI curve, it would be detectable with at least $2\sigma$ confidence.
We note that while a detection of the combined background with O4 sensitivity is unlikely, we start probing the interesting regime if the detectors reach the sensitivity projected for O5.
A detection or nondetection of a stochastic background carries important implications for population models. In the case of a nondetection, certain combinations of parameters governing the merger rate and source properties could be excluded. Nevertheless, because the background depends on numerous parameters and various modeling assumptions, isolating individual constraints remains difficult.

We compare our BBH background forecast, including uncertainties, with recent results in the literature. Compared to Refs.~\cite{LVK-isotropic-o3:2021,GWTC3_population}, we find a significantly broader uncertainty band, particularly at low frequencies. While the uncertainty in the mass distribution is incorporated in a similar way, the broader band is due to the inclusion of uncertainties in the merger rate history. In contrast, the cited works consider only the uncertainty on the local merger rate, assuming a fixed redshift evolution.
Additionally, a minor difference in the spectral shape is introduced by our use of a recent waveform model.
In comparison with Ref.~\cite{Renzini:2024}, our uncertainty range is similar, even slightly narrower. This difference stems from the treatment of merger rate uncertainties: Ref.~\cite{Renzini:2024} varies the low-redshift power-law index [$\gamma$ in~\cref{eq:SFR}] while keeping the high-redshift rate evolution fixed. 
For the background from BNS and NSBH systems, our uncertainty estimates closely match those in Ref.~\cite{GWTC3_population}. This agreement is expected, as the dominant source of uncertainty in these cases is the local merger rate, and we adopt similar assumptions regarding the mass distribution.

\section{Conclusion}
\label{sec:conclusion}

We have presented a comprehensive study of the AGWB produced by CBCs, focusing on contributions from BBH, BNS, and NSBH systems. Starting from fiducial models informed by current observational constraints, we systematically investigated how uncertainties in their mass and spin distributions, as well as different modeling choices for their merger rate histories, affect the resulting energy density spectrum. 
By combining the various uncertainties, we derived credible bands for the GWB from these sources, which span approximately an order of magnitude in fractional energy density.
We found that while local rate uncertainties dominate the uncertainty in the GWB spectrum for BNS and NSBH systems, the merger rate history plays a more significant role for BBHs. In particular, assumptions about the time-delay distribution and metallicity dependence introduce substantial variation in the predicted background. 

Since the completion of this work, the GWTC-4 catalog~\cite{GWTC-4-results:2025} and updated population inferences~\cite{GWTC-4-population:2025} have been released. These updates primarily refine the underlying BBH mass model and tighten the local merger rate estimates. Their impact on the uncertainty in the predicted energy density spectra is expected to be minor---we anticipate slightly narrower uncertainty bands in~\cref{fig:CBCfull} for BBH, BNS, and NSBH systems. However, the dominant source of uncertainty remains the merger rate evolution. 
We note that the forecast in Fig.~7 of Ref.~\cite{Isotropic-O4a:2025}, which incorporates the updated population, reports even larger uncertainties for the BBH background. This difference arises because their approach estimates the low-redshift merger rate evolution directly from observed BBHs rather than using a time-delay and metallicity model, resulting in broader uncertainty compared to our modeling.

Beyond serving as a forecast for potential AGWB detection in upcoming observing runs of ground-based GW detectors, our framework provides a basis for interpreting future measurements. It enables constraints on astrophysical modeling choices and helps in refining our understanding of the compact binary population across cosmic time.
Future extension of this work may incorporate redshift-dependent population models and explore the impact of multiple formation channels.

\begin{acknowledgments}
We would like to thank Irina Dvorkin for helpful comments.
M.E. is supported by the Swiss National Science Foundation Grant Sinergia 213497 and UZH Postdoc Grant, No. [FK-25-109]. We thank the Physik-Institut of the University of Zurich for providing computational resources.

The data that support the findings of this article are openly available at~\cite{data}.
\end{acknowledgments}

\appendix

\bibliography{references}% Produces the bibliography via BibTeX.

\end{document}